\documentclass{article}

\usepackage[utf8]{inputenc} 
\usepackage[T1]{fontenc}    
\usepackage{hyperref}       
\usepackage{url}            
\usepackage{booktabs}       
\usepackage{amsfonts}       
\usepackage{nicefrac}       
\usepackage{microtype}      
\usepackage{graphicx}
\usepackage[nonatbib,preprint]{neurips_2020}
\usepackage{amstext}
\usepackage{amsmath}
\usepackage{verbatim}
\usepackage{graphicx}
\usepackage{subcaption}
\usepackage{algpseudocode}
\usepackage{algorithm}
\usepackage{framed}

\title{Y-Net: Biomedical Image Segmentation and Clustering}

\author{
  Sharmin Pathan \\
  Department of Computer Science\\
  University of Georgia\\
  \texttt{sharmin.pathan25@uga.edu} \\
   \And
   Anant Tripathi \\
   Department of Computer Science \\
   University of Georgia \\
   \texttt{anant.tripathi@uga.edu} \\
   }

\begin{document}

\maketitle

\begin{abstract}
  We propose a deep clustering architecture alongside image segmentation for medical image analysis. The main idea is based on unsupervised learning to cluster images on severity of the disease in the subject’s sample, and this image is then segmented to highlight and outline regions of interest. We start with training an autoencoder on the images for segmentation. The encoder part from the autoencoder branches out to a clustering node and segmentation node. Deep clustering using K-means clustering is performed at the clustering branch and a lightweight model is used for segmentation. Each of the branches use extracted features from the autoencoder. We demonstrate our results on ISIC 2018: Skin Lesion Analysis Towards Melanoma Detection and Cityscapes datasets for segmentation and clustering. The proposed architecture beats U-Net and DeepLab results on the two datasets, and has less than half the number of parameters. We use the deep clustering branch for clustering images into four clusters. Our approach can be applied to work with high complexity datasets of medical imaging for analyzing survival prediction for severe diseases or customizing treatment based on how far the disease has propagated. Clustering patients can help understand how binning should be done on real-valued features to reduce feature sparsity and improve accuracy on classification tasks. The proposed architecture can provide an early diagnosis and reduce human intervention on labeling as it can become quite costly as the datasets grow larger. The main idea is to propose a one-shot approach to segmentation with deep clustering.
\end{abstract}

\section{Introduction}

Several advancements are continuously being made in the field of medical imaging. These range from ways of capturing the complex structures to providing a concrete analysis from captured data. The images are captured in different modalities and algorithms like classification, segmentation, etc can help diagnose patient condition and outline regions of interest like a possible tumor. Over the last decade, longitudinal images are also being increasingly available for studying disease propagation \cite{pathan}. All of this captured image data can help with an early diagnosis for severe diseases whose impact can be reduced in the years to follow. The International Skin Imaging Collaboration (ISIC) 2018 \cite{isic} dataset contains thousands of dermoscopic-images\footnote{ Dermoscopy is an imaging technique that eliminates surface reflection of the skin. By removing surface reflection, visualization of deeper levels of skin is enhanced and results in improved diagnostic accuracy compared to standard photography.} for automated diagnosis of melanoma which is the deadliest form of skin cancer, responsible for an overwhelming majority of skin cancer deaths.

\pagebreak
In our experiments we analyze dermoscopic images of skin lesions from ISIC 2018 dataset, and vegetation cover from Cityscapes \cite{cityscapes} dataset. We use image segmentation to identify the regions of interest in the two datasets, followed by deep clustering to rank lesions based on severity of the disease for ISIC dataset and to cluster images based on the extent of vegetation cover in the Cityscapes dataset. Apart from suggesting if the subject is suffering or not (deep clustering branch), the architecture can also point out the extent to which the disease has propagated (image segmentation branch). Clustering apart from being an unsupervised machine learning algorithm, can also be used to create clusters based on similar features to improve classification accuracy. There are several other applications of unsupervised clustering from pixel data in medical imaging \cite{meyer} demonstrating the flexibility and conceptual power of these techniques. Sharing weights from an autoencoder further helps reduce retraining for multiple branches in the network. With the proposed architecture, we aim to provide a detailed diagnosis by combining clustering and image segmentation results in one-go as an end-to-end process.

Our image segmentation results outperform results from U-Net \cite{ronneberger} and DeepLab \cite{caron} architectures on training the three architectures for just 10 epochs and comparing results. We use K-Means clustering as the target labels to measure performance of our deep clustering branch. Our proposed architecture of image segmentation with deep clustering has total parameters much less than U-Net and DeepLab architectures, while outperforming the results from the two for image segmentation.

\subsection{Related Work}

Our clustering framework for jointly learning the neural network parameters and cluster assignments of the resulting features is based on DeepCluster architecture \cite{caron}. The idea is to iteratively group features using K-means clustering and allow a neural net to learn them to form subsequent cluster assignments under unsupervised learning. This constitutes stage 2 of our pipeline. The main stage being a fully connected convolutional neural network autoencoder which yields precise localization of pixels for segmentation task. The autoencoder is used for image segmentation. We compare our segmentation results with U-Net and DeepLab.

There are approaches for clustering and for segmentation separately. We combine these two and present Y-Net. Y-Net architecture gives a better classification accuracy and is proposed to be used for tasks not limited to classification but analyzing the degree of disease propagation. It then also provides basis to this analysis by pixel localization of the regions of interest in the image that help define the extent of how adversely the patient has been affected.

\section{Network Architecture}

\begin{figure}[t]

  \centering
  \fbox{ \includegraphics[scale=0.7]{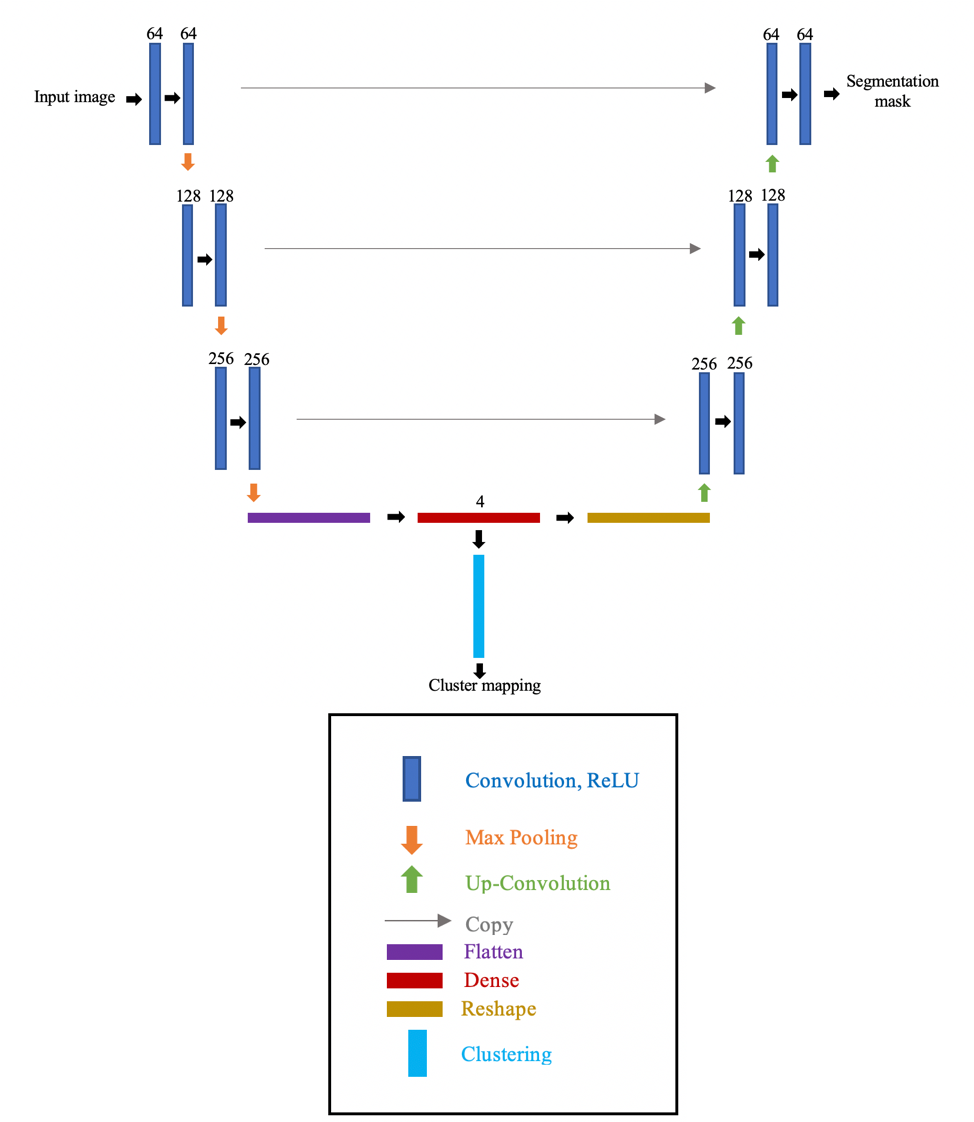}} 
  \newline
  \caption{Y-Net architecture. Each blue bar corresponds to a multi-channel feature map. The number of channels is denoted on top of each bar. Dense layer has 4 output units which further branches out to clustering and maps features to 4 clusters.}
  \label{network}
\end{figure}

Figure \ref{network} illustrates our proposed network architecture. The autoencoder is used for precise image segmentation detailing regions of interest in the image that define severity. We train this autoencoder and use weights from the encoder and add another branch to do clustering. This branch is created by adding a clustering layer to the encoder. The architecture forms a ‘Y’ like structure. Following subsections describe the architecture in more detail. The dense layer that branches out to clustering can be set to the required number of clusters. For our experiments, we used dense layers with 4 units. Our entire architecture (segmentation with clustering) has around 20 million parameters which is nearly half the number of parameters compared to U-net or DeepLab Xception net.

\subsection{Autoencoder}

This pathway in the network accepts baseline image of a subject and extracts features that will be used in clustering as well as segmentation. Convolutional Neural Networks are a popular choice for mapping raw images to a vector space of fixed dimensionality. We use a fully connected convolutional neural network for the autoencoder. The encoder path uses repeated application of two 3x3 convolutions, each followed by a rectified linear unit (ReLU) and a 2x2 max pooling operation with stride 2 for downsampling. At each downsampling step we double the number of feature channels. We then flatten the extracted features and pass it through a dense layer with 4 units which can be used to extend the clustering branch and map the features to 4 clusters. Every step in the decoder path consists of an upsampling of the feature map followed by a 2x2 convolution that halves the number of feature channels, a concatenation with the correspondingly cropped feature map and two 3x3 convolutions, each followed by ReLU. The required input image dimensions are 512x512. The autoencoder is trained for image segmentation. The total parameters for our autoencoder are 15,510,917 which is less than half the number of parameters of U-Net or DeepLab and our network still outperforms their results.

\subsection{Deep Clustering}

We extract the pretrained encoder from our autoencoder. The encoder architecture has its last dense layer with 4 output neurons that extract features from each image and map it to our 4 clusters. We then branch out to a clustering layer from the encoder using k-means to generate cluster centroids around these 4 outputs. By training the autoencoder, we have its encoder part learned to compress each image into four floating point values. We build a custom clustering layer to convert input features to cluster label probability. The probability is calculated to measure similarity between an embedded point and a centroid. The clustering layer’s weights represent cluster centroids. They are grabbed by fitting k-means on the encoder’s prediction. The clustering layer is trained to act similar to K-means for clustering. We fine-tune the encoder weights to improve cluster assignment and feature representation simultaneously. For this purpose we deployed a centroid-based target probability distribution and tried to minimize its KL divergence against model clustering results.

\section{Experiments}

In our experiments we analyze dermoscopic images of skin lesions from ISIC 2018 dataset, and vegetation cover from Cityscapes dataset. We train our autoencoder, U-Net, and DeepLab architecture with Xception Net for 10 epochs with a batch size of 1 and compare results on both the datasets. Table \ref{iou} shows IOU scores using Y-Net autoencoder, U-Net, and DeepLab Xception Net on ISIC and Cityscapes datasets. In our architecture, we use Adam optimization and a dropout rate of 0.5. The input images and target masks were normalized and resized to 512x512 to be fed to the network. We then fine tune the weights when we retrain with additional clustering layers following the encoder for another 10 epochs. 

\begin{table}[!htb]
  \caption{IOU scores of Y-Net autoencoder, U-net, and DeepLab Xception Net on the test set. The three models were trained for only 10 epochs. Training for further epochs improves performance. \newline}
  \centering
  \begin{tabular}{llll}
    \toprule
         Dataset & Y-Net autoencoder     & U-Net     & DeepLab Xception Net  \\
    \midrule
    ISIC 2018 melanoma dataset & \multicolumn{1}{r}{0.6225} & \multicolumn{1}{r}{0.5334} & \multicolumn{1}{r}{0.4892}      \\
    Cityscapes Vegetation cover & \multicolumn{1}{r}{0.7632} & \multicolumn{1}{r}{0.7066} & \multicolumn{1}{r}{0.7118}         \\
    \bottomrule
  \end{tabular}
  \label{iou}
\end{table}

\begin{figure}[!htb]
  \begin{framed}
  \begin{subfigure} [t] {.5\textwidth}
  \includegraphics[scale=0.3]{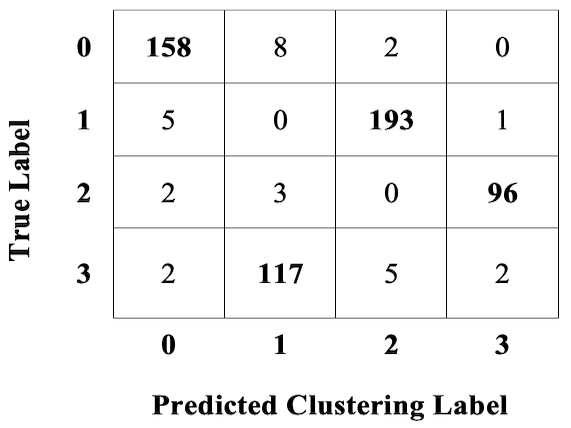}
  \caption{Clustering confusion matrix on ISIC dataset}
  \end{subfigure}
  \hfill
  \begin{subfigure} [t] {.5\textwidth}
  \includegraphics[scale=0.3]{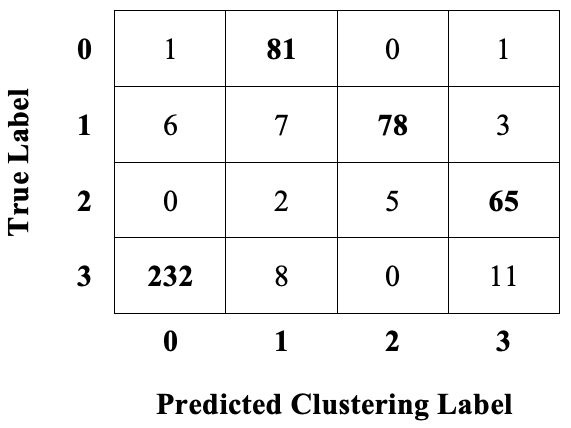} 
  \caption{Clustering confusion matrix on cityscapes dataset}
  \end{subfigure}  
  \end{framed}
  \caption{Confusion matrix for Y-net clustering on ISIC 2018 melanoma test set, and on cityscapes test set. For ISIC test set, our clustering label 0 maps to true label 0, clustering label 1 maps to true label 3, clustering label 2 maps to true label 1, and clustering label 3 maps to true label 2. For cityscapes vegetation cover in the test set, clustering label 0 maps to true label 3, clustering label 1 maps to true label 0, clustering label 2 maps to true label 1, and clustering label 3 maps to true label 2.}
  \label{confusion_matrix}
  \end{figure}

\begin{figure}[!htb]
\begin{framed}
  
  \begin{subfigure} [t] {.3\textwidth}
  \centering
  \includegraphics[scale=0.18]{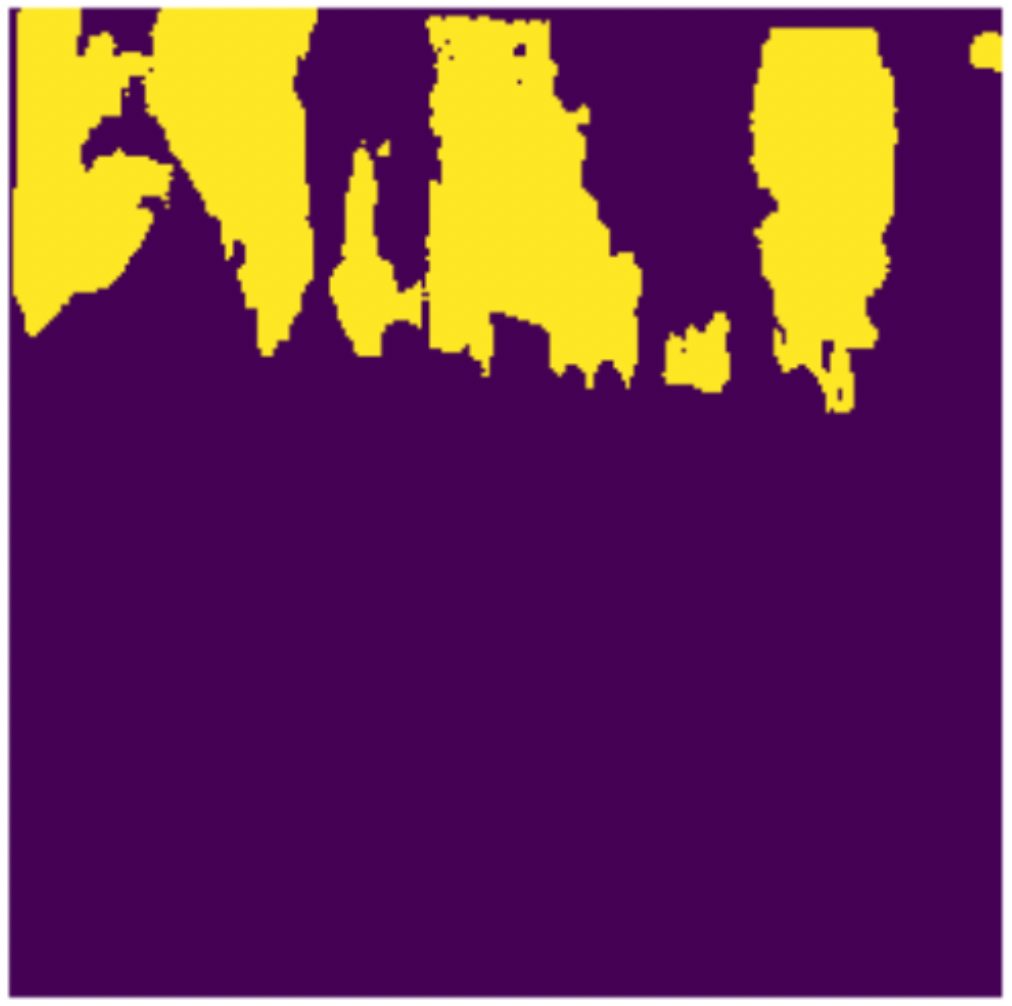}
  \caption{Y-net mask prediction}
  \end{subfigure}
  \hfill
  \begin{subfigure} [t] {.3\textwidth}
  \centering
  \includegraphics[scale=0.18]{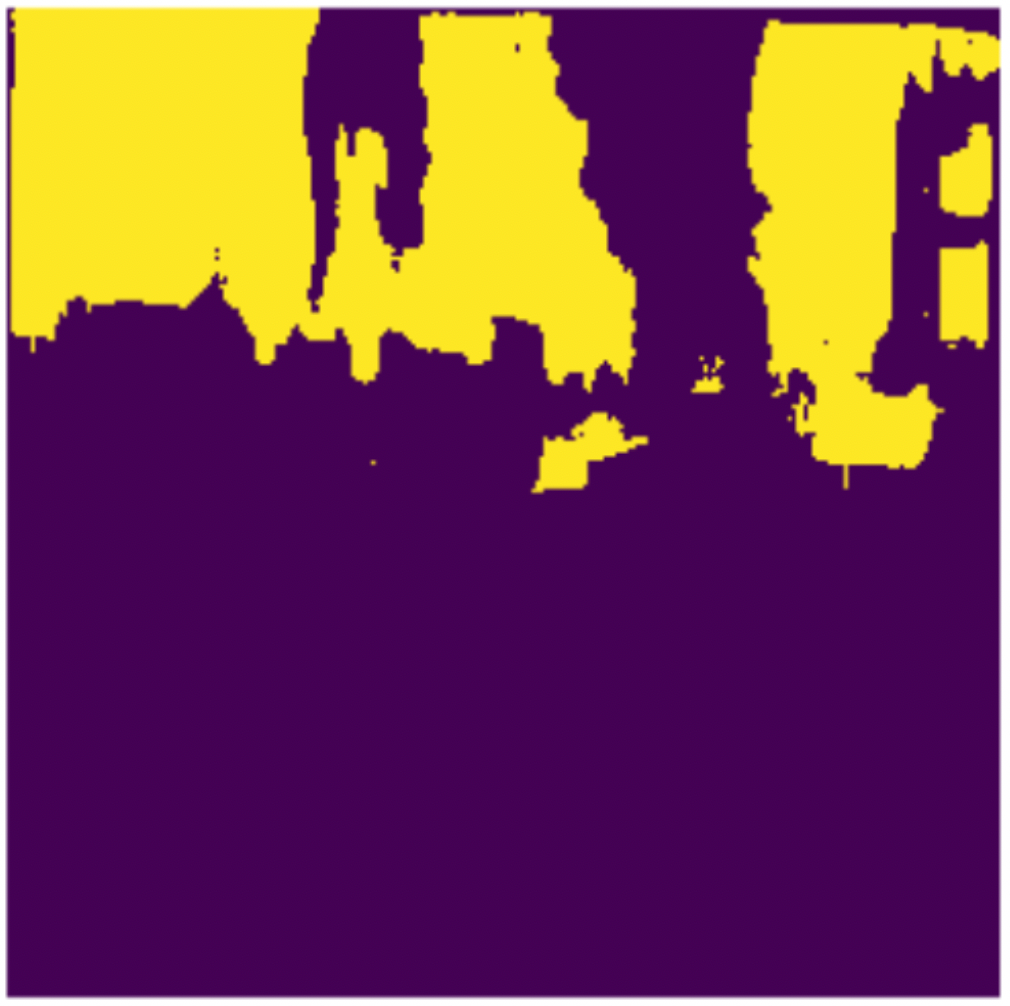} 
  \caption{U-net mask prediction}
  \end{subfigure}
  \hfill
  \begin{subfigure} [t] {.3\textwidth}
  \centering
  \includegraphics[scale=0.18]{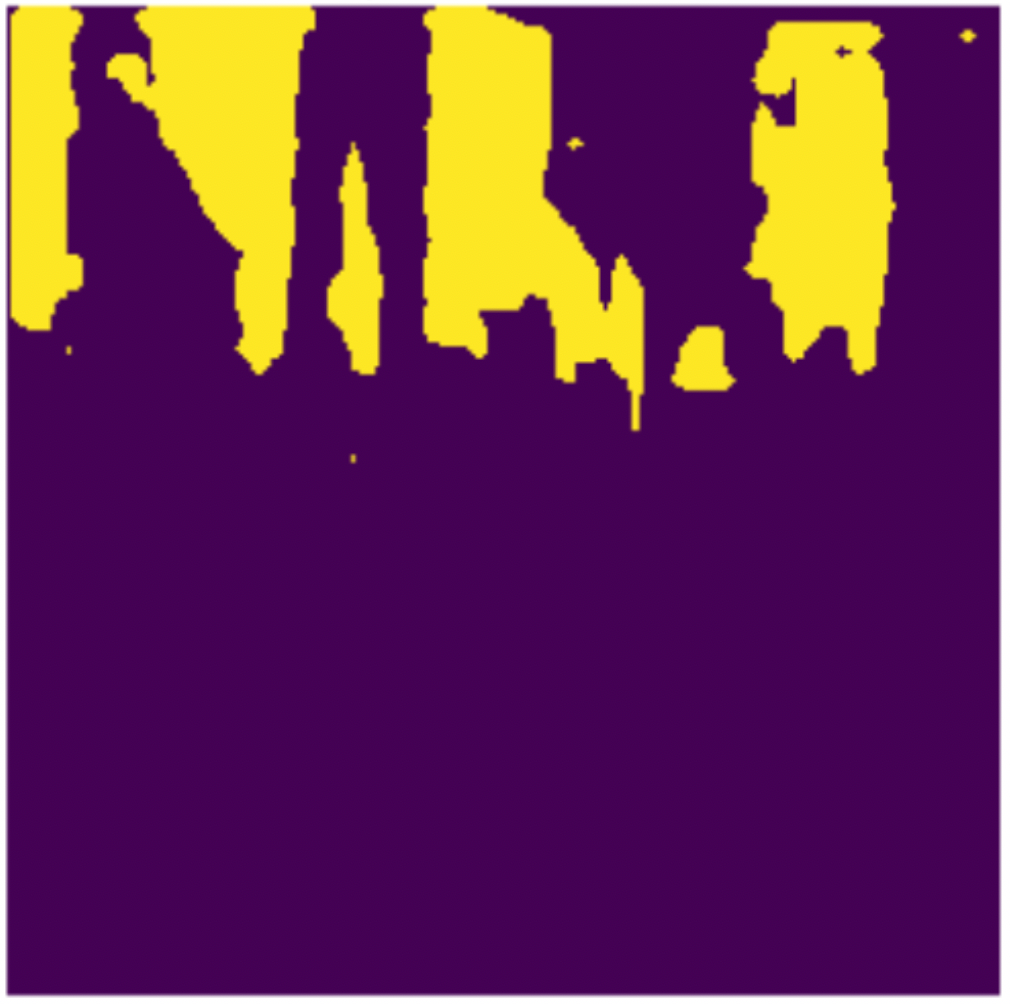} 
  \caption{DeepLab mask prediction \newline}
  \end{subfigure} 
  \newline
  \newline
  \begin{subfigure} [t] {.3\textwidth}
  \centering
  \includegraphics[scale=0.18]{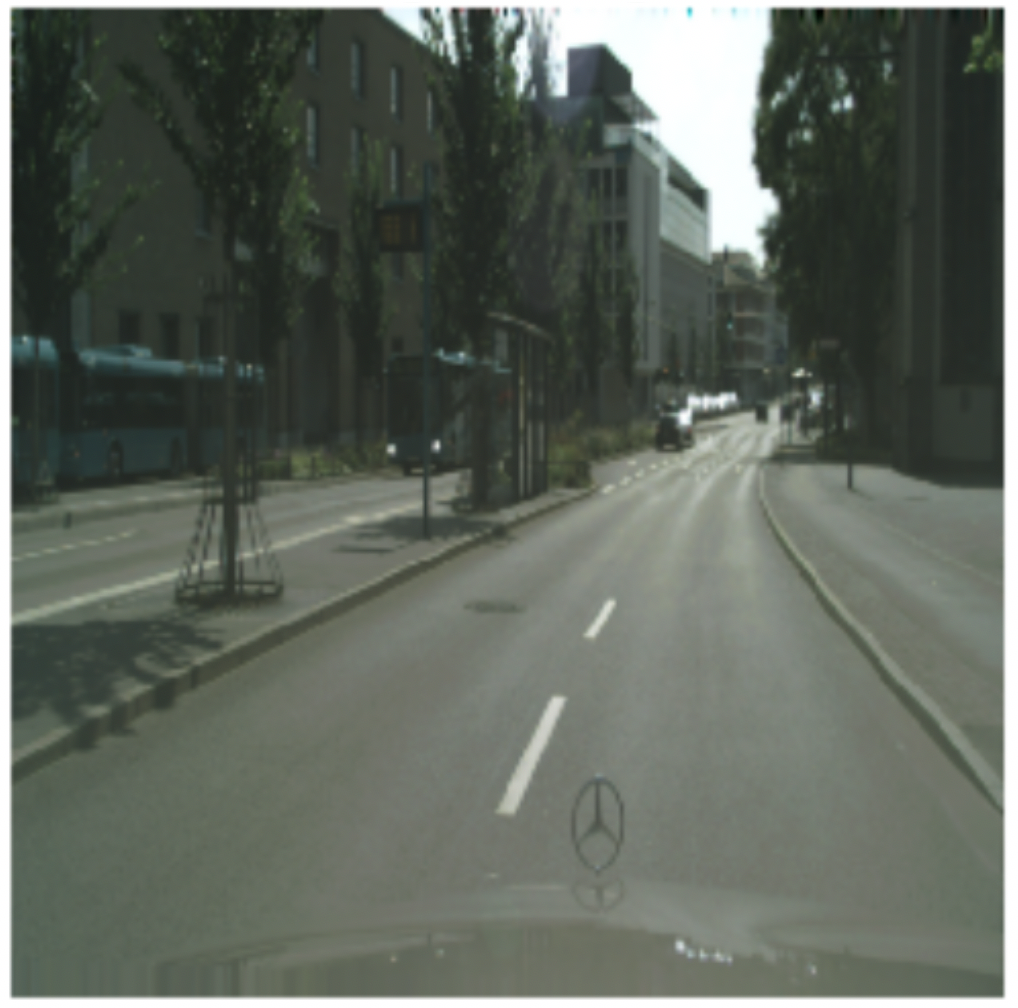} 
  \caption{Original image}
  \end{subfigure} %
  \hspace{0.5cm}
   \begin{subfigure} [t] {.3\textwidth}
   \centering
  \includegraphics[scale=0.18]{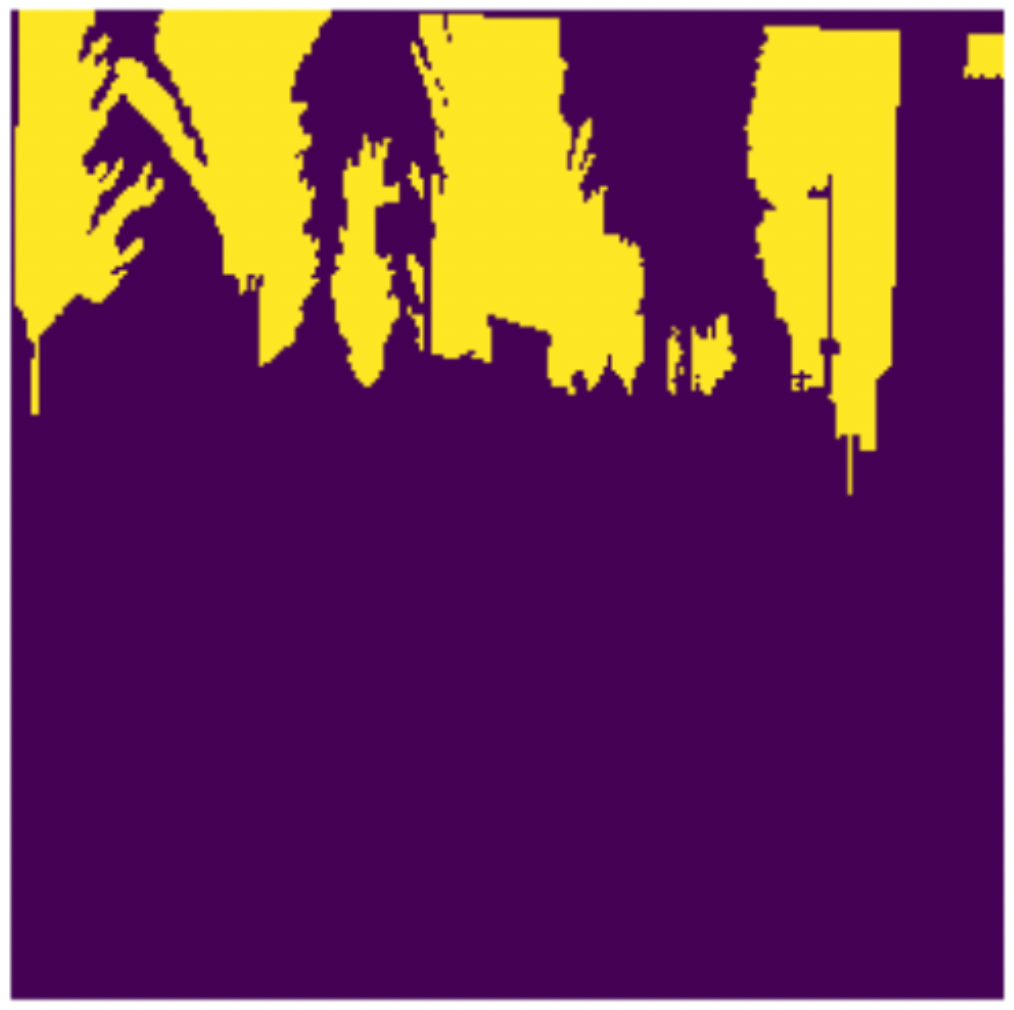} 
  \caption{Ground truth mask}
  \end{subfigure}%
  \hfill
   \end{framed}
  \caption{Vegetation cover on a sample image from cityscapes dataset. The bottom row contains original input image and its ground truth mask. The top row shows results from Y-Net, U-Net, and DeepLab Xception net predictions for the same input image. \newline }
  \label{cityscapes}
 
\end{figure}

\begin{figure}[!htb]
\begin{framed}
  \begin{subfigure} [t] {.3\textwidth}
  \centering
  \includegraphics[scale=0.18]{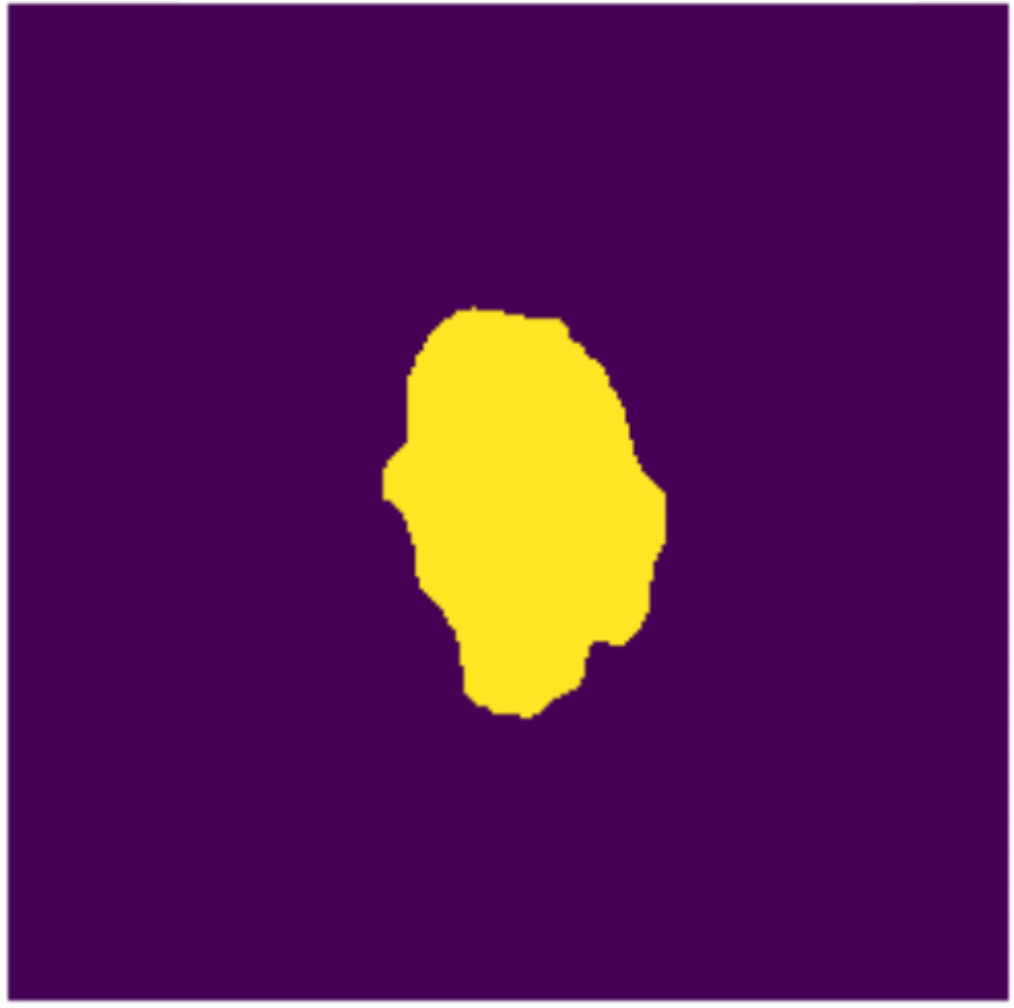}
  \caption{Y-net mask prediction}
  \end{subfigure}
  \hfill
  \begin{subfigure} [t] {.3\textwidth}
  \centering
  \includegraphics[scale=0.18]{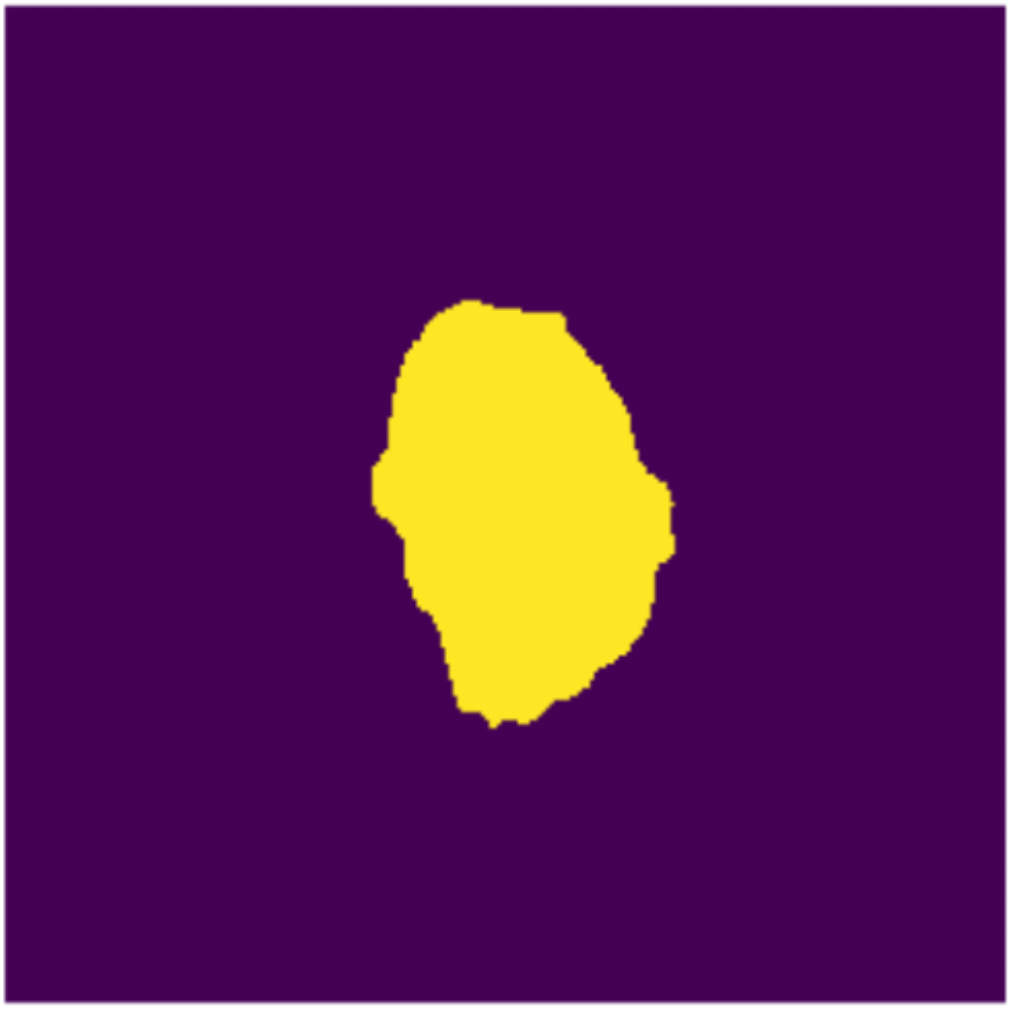} 
  \caption{U-net mask prediction}
  \end{subfigure}
  \hfill
  \begin{subfigure} [t] {.3\textwidth}
  \centering
  \includegraphics[scale=0.18]{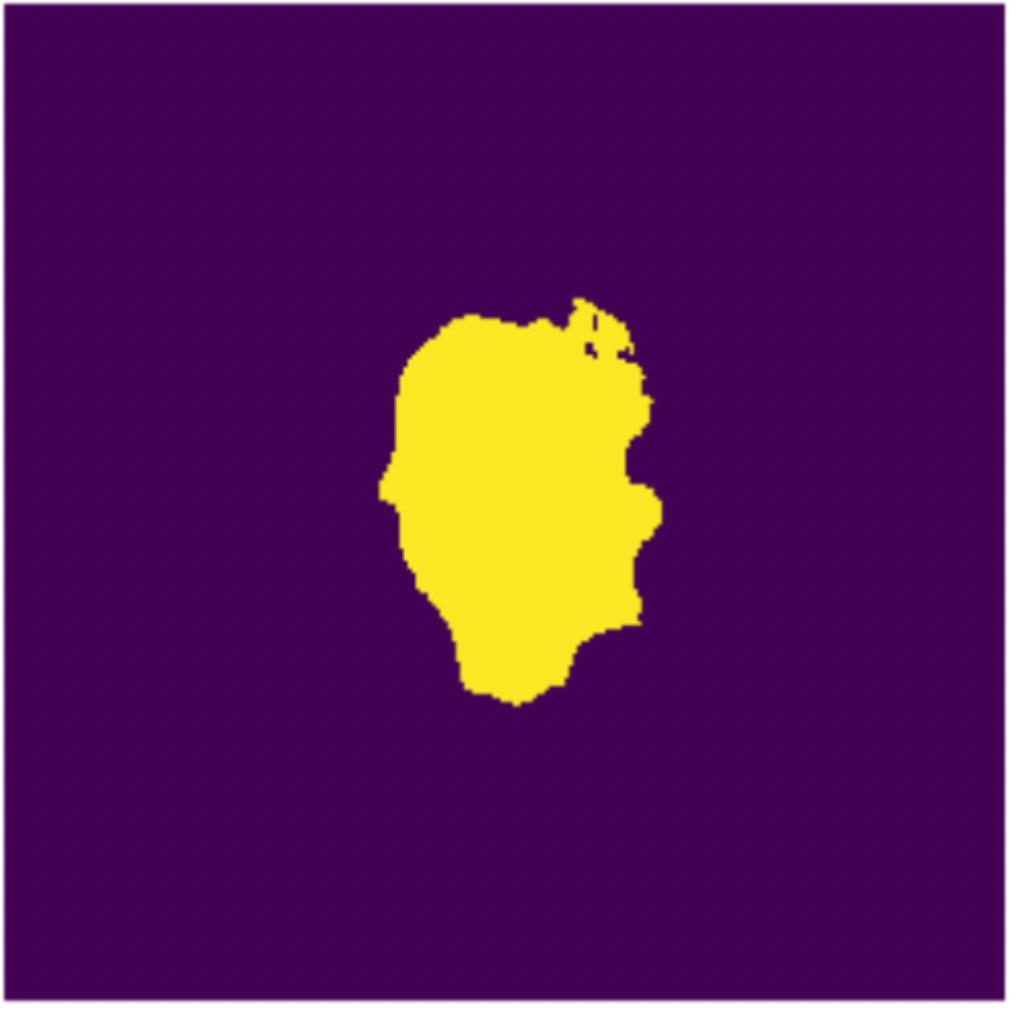} 
  \caption{DeepLab mask prediction \newline}
  \end{subfigure} 
  \newline
  \newline
  \begin{subfigure} [t] {.3\textwidth}
  \centering
  \includegraphics[scale=0.18]{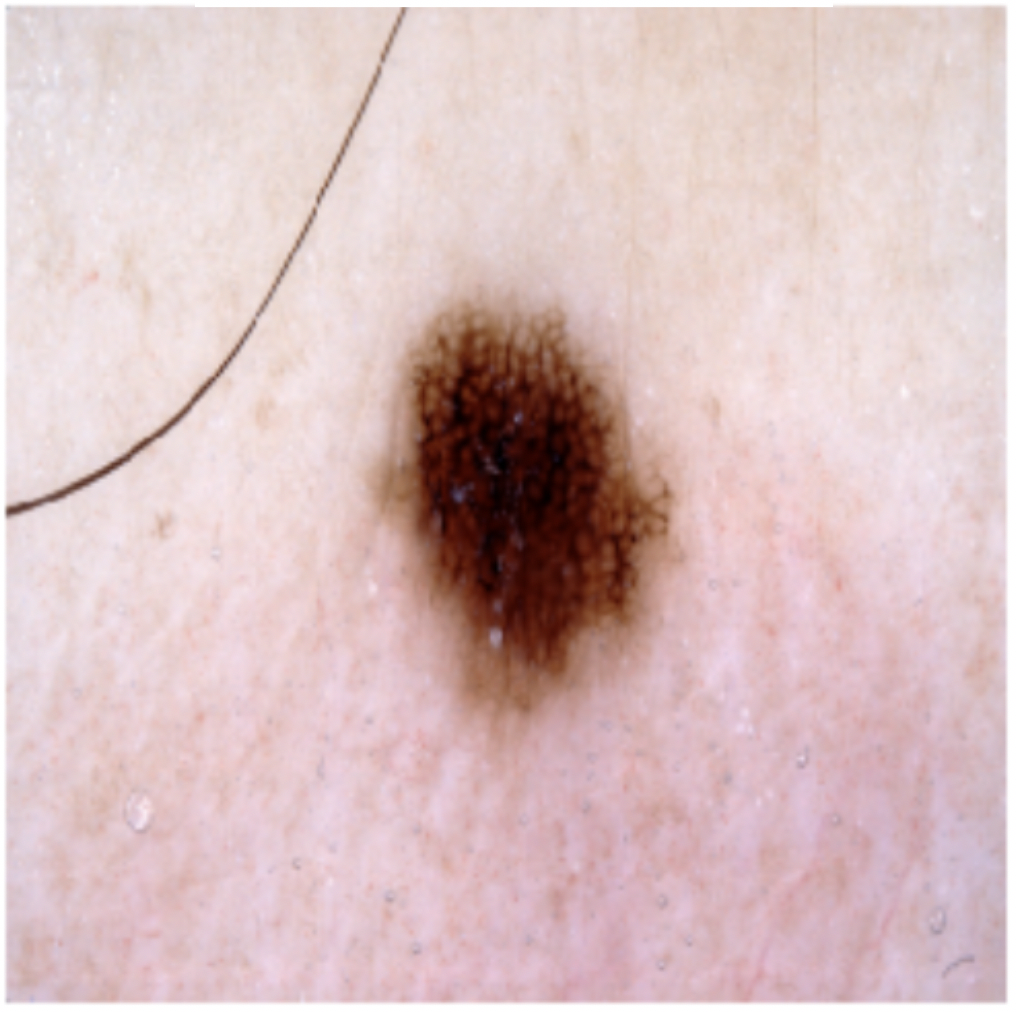} 
  \caption{Original image}
  \end{subfigure} %
  \hspace{0.5cm}
   \begin{subfigure} [t] {.3\textwidth}
   \centering
  \includegraphics[scale=0.18]{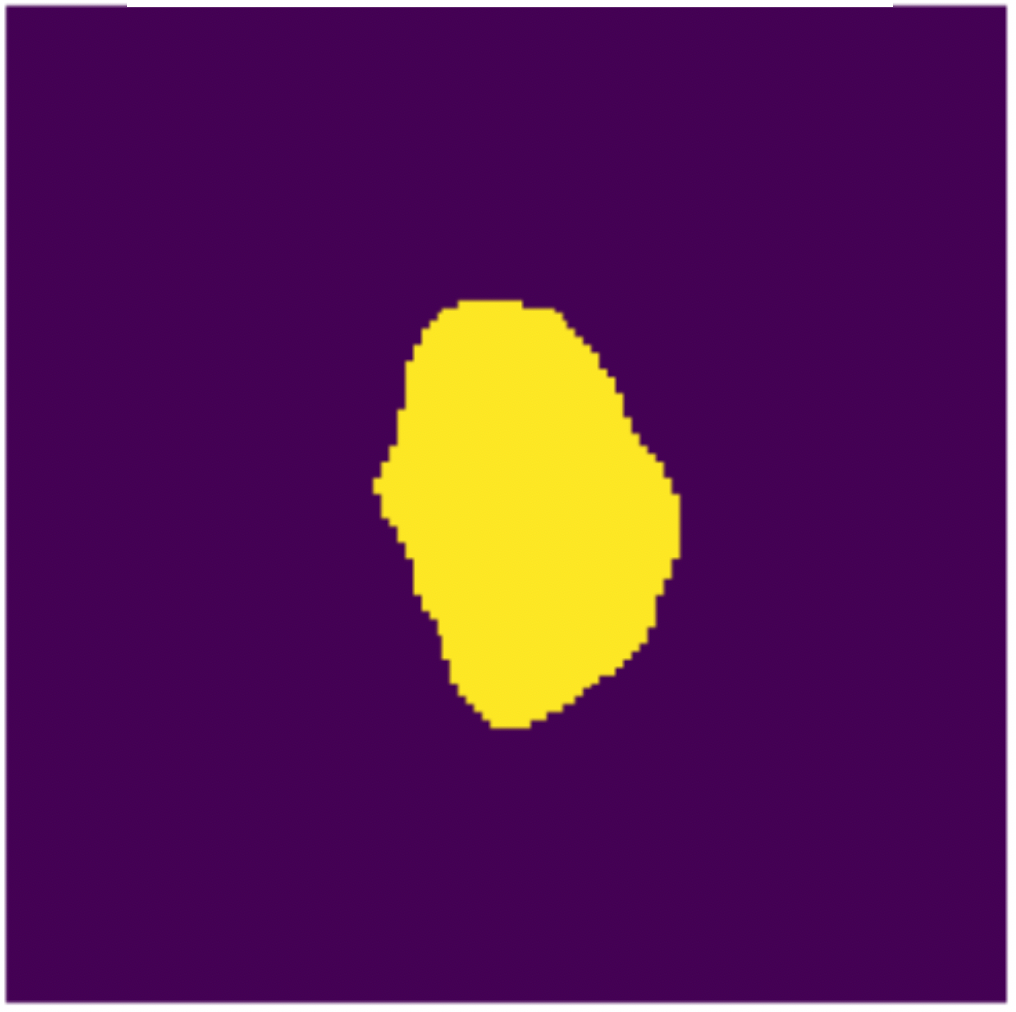} 
  \caption{Ground truth mask}
  \end{subfigure}%
  \hfill
   \end{framed}
  \caption{The bottom row contains an original dermoscopic image from ISIC 2018 melanoma dataset and its ground truth mask. The top row shows results from Y-Net, U-Net, and DeepLab Xception net predictions on the original image. \newline }
  \label{isic}
\end{figure}

Figure \ref{cityscapes} shows our results on cityscapes dataset, and Figure \ref{isic} shows our results on ISIC 2018 melanoma dataset. Since the two datasets didn't have clustering labels, we make predictions using k-means and use it as our target labels for the clustering model. We iteratively refine clusters by learning from the high-confidence assignments with the help of auxiliary target distribution and clustering output. We use KL divergence as the loss function for our clustering branch to minimize it so that the target distribution is as close to the clustering output distribution as possible. Figure \ref{confusion_matrix} shows the clustering confusion matrices for the two datasets.

\pagebreak 

\pagebreak
\section*{Conclusion}

We demonstrate that our light-weight model achieves good performance with unsupervised clustering and helps detail the level of severity of a disease alongside image segmentation. This approach can also help with early diagnosis of a disease. We demonstrate a one-shot approach for image-segmentation with deep clustering. The model inherits good properties of image segmentation from the autoencoder with weight sharing across deep clustering. The approach can easily be extended to several other applications like defect detection in manufacturing where different types of defects need to be identified in the part being manufactured, or as demonstrated it can be used for disease diagnosis from biomedical images. We implemented our architecture to support 2D images; however, it can be straightforwardly extended for 3D images. One possible challenge in 3D implementation is a shortage of GPU memory, which could be solved by using 3D image patches instead of the whole 3D volume.



\small

\begin{thebibliography}{1}
\bibitem{pathan} Pathan S, Hong Y, 2018. Predictive Image Regression For Longitudinal Studies with Missing Data, Conference on Medical Imaging with Deep Learning. {\it arXiv preprint arXiv:1808.07553}.

\bibitem{isic} ISIC 2018: Skin lesion analysis towards melanoma detection. \newline \url{https://challenge2018.isic-archive.com}.

\bibitem{noel} Codella N, Rotemberg V, Tschandl P, Celebi M E, Dusza S, Gutman D, Helba B, Kalloo A, Liopyris K, Marchetti M, Kittler H, Halpem A, 2018. Skin Lesion Analysis Toward Melanoma Detection: A Challenge Hosted by the International Skin Imaging Collaboration (ISIC). {\it arXiv:1902.03368}.

\bibitem{tschandl} Tschandl P, Rosendahl C, Kittler H, 2018, The HAM10000 dataset, a large collection of multi-source dermatoscopic images of common pigmented skin lesions. {\it Sci. Data 5, 180161 doi:10.1038/sdata.2018.161}.

\bibitem{cityscapes} Cityscapes dataset. \url{https://www.cityscapes-dataset.com}.

\bibitem{cordts} Cordts M, Omran M, Ramos S, Rehfeld T, Enzweiler M, Benenson R, Franke U, Roth S, and Schiele B, 2016. The Cityscapes Dataset for Semantic Urban Scene Understanding, in Proc of the IEEE Conference on Computer Vision and Pattern Recognition (CVPR).

\bibitem{meyer} Meyer-Baese A, Theis FJ, Gruber P, Wismueller A, Ritter H, 2005, Application of Unsupervised Clustering Methods to Medical Imaging. WSOM.

\bibitem{ronneberger} Ronneberger O, Fischer P, Brox T, 2015, U-Net Convolutional Networks for Biomedical Image Segmentation. MICCAI.

\bibitem{caron} Caron M, Bojanowski P, Joulin A, and Douze M, 2005. Deep Clustering for Unsupervised Learning of Visual Features. {\it CoRR: abs-1807-05520}.

\end{thebibliography}
\end{document}